\documentclass[a4paper,10pt]{article}
\usepackage{graphicx} % Required for inserting images
\usepackage[lmargin=1.75cm, rmargin=1.75cm, bmargin=2cm, tmargin=2cm]{geometry}
\usepackage{subcaption}
\usepackage{float}
\usepackage{svg}
\usepackage[square,sort,comma,numbers]{natbib}

\title{Estimating Hydrodynamic Coefficients for Floating Offshore Structures from Movement Data Using Physics-Informed Neural Networks}
\author{Anders Schou, Jens Visbech, and Allan Peter Engsig-Karup*\\ \\
Department of Applied Mathematics and Computer Science, \\Technical University of Denmark, 2800, Kgs. Lyngby, Denmark\\ \\
\small{*Corresponding author email: apek@dtu.dk}}
\date{July 2026}

\usepackage{amsmath, amsfonts, amssymb, amsthm}
\usepackage{tikz}
\usepackage{xcolor}

\begin{document}
\maketitle

\section*{ABSTRACT}
We present a method for estimating the hydrodynamic coefficients in the Cummins equations using time-series data from a moving body, such as a floating offshore structure. The proposed data-driven method is based on incorporating the Cummins equations governing the dynamics of a structural body interacting with water waves into a physics-informed neural network (PINN), along with available motion data. The proposed method first estimates the structure's state in terms of translational and rotational degrees of freedom, and then solves the inverse problem to determine the hydrodynamic forces acting on the body, expressed in terms of added mass, damping coefficients, and/or hydrostatic restoring. The Cummins equations are formulated as a first-order system, and both state and parameter estimation are performed using PINNs. The method is verified on the free decay of a sphere and a box. The results demonstrate that it is possible to estimate the state and hydrodynamic coefficients accurately, although accuracy depends on the volume and quality of the movement data. \\ \\

\noindent \textbf{Keywords:}\\
Neural networks, physics-informed neural networks, hydrodynamic coefficients, impulse response function, parameter estimation, Cummins equations, wave-structure interactions, physics-informed machine learning, scientific machine learning.
\section{INTRODUCTION}

% General intro and motivation for solving the problem
A central challenge in ocean and marine engineering is simulating wave-structure interactions (WSIs) with floating objects such as ships, floating offshore wind turbines, and wave energy converters. Therefore, when designing such structures, it is of great interest to simulate their motion behavior in various sea states. This can support the expensive and time-consuming model and full-scale testing phase in physical wave tanks or the ocean. Within this engineering field, the structural movements (translatory and rotational) can often be described from a linear perspective derived from Newton's second law and linear potential flow theory, given an assumption of small wave height and structural movements compared to the wavelength; ultimately, arriving at the well-known Cummings equations \cite{cummins1962impulse}. These equations describe the time-domain motion of structures such as floating bodies, based on their geometric characteristics and the forces exerted on them.  The equations are formulated in terms of linearised hydrodynamic coefficients for mass, added mass, damping, hydrodynamic restoring, etc. If these coefficients are known, it is possible to solve the forward problem of simulating the behavior of the moving structure given the loads on the structure \cite{MortensenEtAl2021}. However, if these are not correct or only partially known, which is often the case for more complex structures, they need to be estimated to be able to accurately model the structural movements.
The task of computing estimates of such coefficients involves solving multiple boundary-value problems in the frequency domain using BEM, as can be done with state-of-the-art commercial software such as WAMIT \cite{wamit} or open-source software such as NEMOH \cite{KURNIA2023108885}. Also, equivalently, a single time-dependent impulsive problem can be solved in the time domain, from which the coefficients can be linked to the frequency domain through Fourier transforms, e.g., see \cite{visbech2024solving, liapis1986time}. % [TODO: add indirect measurements refs]

In design and development loops, engineers often optimize for performance, e.g., maximizing power extraction in a wave energy converter design. This optimization process is closely linked to the structure's geometrical shape and layout, and the design slowly converges to its final state. Here, minor adjustments might be needed, such as increasing the heave plate diameter of a floating offshore wind turbine foundation or reducing the draft of a new ship design. Here, physical models- and/or full-scale models are presumably already built. Instead of running numerical simulations for the latest design, one can make the actual physical adjustments to the structure, perform movement tests (free decay and/or wave exposure), and use this data to predict the hydrodynamic coefficients. With this, the hydrodynamic coefficients can be estimated from available sensor/movement data without solving a series of time-dependent boundary value problems or setting up a numerical solver, as described in \cite{MortensenEtAl2021}.

% Neural networks gaining attention
Recently, neural networks have gained renewed interest as surrogate models for boundary-value problems or as a means of setting up numerical solvers to do so, due to their flexibility and approximation capabilities. In recent years, an emerging field in data science, referred to as {\em scientific machine learning} (SciML) \cite{baker2019workshop}, may be viewed as the intersection of well-established data science and scientific computing fields. SciML focuses on researching new computing methods that combine measurement data with domain knowledge expressed as differential equations to meet first-principles modeling constraints, helping to establish the validity of predictions and providing a basis for generalization and interpretability. Recent reviews are given in \cite{Karniadakis2021,DiColaEtAl2022}. Modeling frameworks such as the physics-informed neural network (PINN) \cite{raissi2019physics} have paved the way for alternative methods of solving inverse problems in science and engineering. Recently, neural operators have also been used to estimate the motion of floating structures from time-series sea-state data \cite{cao2024deep}.

% Approach and contributions in this paper
In this work, we propose a data-driven approach to approximating hydrodynamic coefficients from sensor-measured movement data. The approach is based on the PINN methodology \cite{raissi2019physics} and is inspired by the universal PINN (UPINN) \cite{podina2023universal} framework that conceptualizes the incorporation of neural networks that account for incomplete parts of a model to recover missing parts of a model to make a best fit with available (often noisy) measurement data. The quality of the data used will affect the method's accuracy. The purpose is not to outperform existing methods in terms of accuracy; rather, the benefits of the proposed method lie in its ease of application in practice. 

\subsection{Contributions}
To summarise the contributions of our work presented in this study:
(i) Based on known frameworks, we propose an alternative method of estimating hydrodynamic coefficients from a data-driven perspective using PINNs by incorporating Cummins equations as a physics loss term. (ii) We apply state-of-the-art techniques to facilitate faster training and more accurate estimates. (iii) We test the proposed method on two types of floating structures in free decay. (iv) Based on the results, we postulate that the proposed method will be successful in more advanced WSI cases, which is left for future work to validate.

\section{PHYSICS-INFORMED MACHINE LEARNING}
\subsection{Governing Equation}
The Cummins equations govern linearised wave-structure interactions,
\begin{equation}
    \left(m_{jk}+a_{jk}^\infty\right)\ddot{x}_{k}(t)+\int_0^tK_{jk}(t-\tau)\dot{x}_{k}(\tau)d\tau+c_{jk}x_{k}(t)=F_{j}(t),\label{eq:Cummins}
\end{equation} 
where $j=1,\dots,6$ and we implicitly sum over all degrees of freedom $k=1,\dots,6$ of the movement $x_k$. Surge, sway, and heave are denoted by $k = 1,2,3$ and roll, pitch, and yaw by $k = 4,5,6$, respectively. The quantities $m_{jk}$, $a_{jk}^\infty$, and $c_{jk}$ are the coefficients of (fixed) mass, infinite added mass, and hydrostatic restoring, respectively. Additionally, the impulse response function (IRF), denoted by $K_{jk}(t)$, and the external forces $F_{j}(t)$ influence the dynamics. The IRF is related to the frequency-dependent added mass and damping coefficients through the relations,
\begin{subequations}
\begin{align}
    K_{jk}(t)&=-\frac{2}{\pi}\int^\infty_0 \omega \left(a_{jk}(\omega)-a^\infty_{jk}\right)\sin(\omega t) d\omega\label{eq:IRF2A}\\
    &=\frac{2}{\pi}\int^\infty_0 b_{jk}(\omega)\cos(\omega t) d\omega\label{eq:IRF2B}.
\end{align}
\end{subequations}
The general goal is to determine the coefficients $a_{jk}(\omega), b_{jk}(\omega), a_{jk}^\infty, c_{jk}$, which is typically done by solving a series of boundary values problem in the frequency domain or a time-dependent initial boundary value problem in the time domain, \cite{wamit,liapis1986time,visbech2024solving}.

\subsection{Physics-Informed Neural Networks}
Neural networks have been applied across many areas of data-driven science and engineering due to their high flexibility and ability to approximate a wide range of functions \cite{cybenko1989approximation, hornik1989multilayer}. The most basic form of a neural network is the feed-forward neural network, also known as a multilayer perceptron (MLP). The MLP takes an input vector ${\bf x}=(x_1,x_2,...,x_d)$ of size $d$ and does a composition of multiple layers $h_\ell$ defined as
\begin{equation}
    h_\ell({\bf x})=\sigma_\ell \left({\bf W}_\ell {\bf x}+{\bf b}_\ell\right),
\end{equation}
where for a given layer ${\bf W}_\ell\in\mathbb{R}^{m_l \times d_l}$ is a weight matrix, ${\bf b}_\ell\in\mathbb{R}^{m_l}$ is a bias vector, and $\sigma_\ell$ is an element-wise nonlinearity called the activation function. In the last layer of the network, the activation is generally omitted to avoid limiting the network's expressivity to the activation's codomain. When training the MLP, an objective function (typically called the loss function) is minimized with respect to the parameters $\Theta_h=\left\{{\bf W}_\ell,{\bf b}_\ell\right\}_{\ell=1}^{L}$. This way, a total of $L$ {\em layers} can be defined to form a compositional function that transforms an input vector
\begin{align*}
f_{\theta}({\bf x}) = \sigma_L({\bf W}_L(\sigma_2({\bf W}_2 \sigma_1({\bf W}_1 {\bf x}+{\bf b}_1) + {\bf b}_2)\cdots)+{\bf b}_L), \quad \forall {\bf x}\in D,
\end{align*}
where ${\bf x}\in D \subset \mathbb{R}^d$, where $D$ is a domain of dimension $d$ which we may refer to as the {\em parameter space}.

Especially in the last decade, with the development of deep learning frameworks such as PyTorch \cite{paszke2019pytorch}, which can utilize graphics processing units (GPUs) and calculate gradients using automatic differentiation (AD) via backpropagation \cite{Rumelhart1986LearningErrors}, training of neural networks to determine the parameters has become much more accessible, leading to an immense amount of research in their applications.

%\subsubsection*{Physics-Informed Neural Networks}
The idea of utilizing physics-informed neural networks (PINNs) originates from works such as \cite{lee1990neural,psichogios1992hybrid,dissanayake1994neural,lagaris1998artificial}, and this idea was later popularized by Raissi et al.  \cite{raissi2019physics} and later given support via an open source library DeepXDE \cite{lu2021deepxde}, and have since then quickly gained popularity and have been used across computational science and engineering. The general idea of the PINN is to represent the solution to a differential equation as a neural network. Following recent advances in deep learning frameworks, the further development of this original concept of using model equations in the loss function has seen strong adoption, driven by interest in leveraging both data and model equations to improve the interpretability and reliability of data-driven approaches. To construct a PINN, suppose we set up the network to be differentiable with respect to the input variables. In that case, we can use AD to compute the derivatives in the equation and incorporate the equation residuals into a loss function. The framework can be applied to many types of equations, e.g., ordinary differential equations (ODEs) and partial differential equations (PDEs).\\ \\
Consider the general problem,
\begin{subequations}
\begin{align}
    \mathcal{N}(u)&=f(\boldsymbol{x}),&\boldsymbol{x}&\in\Omega,\label{eq:diffeq}\\
    \mathcal{B}(u)&=g(\boldsymbol{x}),&\boldsymbol{x}&\in\partial\Omega,\label{eq:init_bound_cond}
\end{align}
\end{subequations}
where $\Omega\subset\mathbb{R}^d$ in general is a spatiotemporal domain, $\mathcal{N}$ is some linear or nonlinear differential operator and the operator $\mathcal{B}$ denotes the initial and/or boundary conditions. A function $u:\Omega\rightarrow\mathbb{R}$ is a solution to the problem if it satisfies the equations \eqref{eq:diffeq}-\eqref{eq:init_bound_cond}. By using a neural network as a surrogate $\hat{u}$ parametrized by $\Theta_u$, we can formulate a physics-informed loss function by incorporating the differential equations as
\begin{equation}
    \mathcal{L}(\Theta_u)= \lambda_1\left(\mathcal{N}(\hat{u})-f(x)\right)^2 + \lambda_2\left(\mathcal{B}(\hat{u})-g(x)\right)^2,\label{eq:pinn_loss}
\end{equation}
where $\lambda_1$ and $\lambda_2$ are positive real-valued parameters determining the contribution of each loss term. It is clear that the minimum of $\mathcal{L}$ is 0, and it is 0 if and only if $\hat{u}$ satisfies the equations \eqref{eq:diffeq}-\eqref{eq:init_bound_cond}. Thus, minimizing the loss function with respect to $\Theta_u$, the surrogate $\hat{u}$ can approximate the true solution $u$. The loss function \eqref{eq:pinn_loss} can be extended to a system of differential equations straightforwardly by including a loss term for each equation in the system.

The central challenge of PINNs lies in the difficulty of the multi-objective optimization problem. Many extensions to the PINN framework have been proposed to facilitate faster convergence and higher surrogate accuracy. One is the self-adaptive PINN (SA-PINN) \cite{mclenny2023selfadaptive}. The SA-PINN assigns trainable weights to each data point and/or residual point in the problem, placing greater emphasis on points with high loss.

The PINN framework assumes that the equations are known (see \eqref{eq:diffeq}-\eqref{eq:init_bound_cond}); however, in some cases it may be of interest to model unknown terms. An example of this could be movement-dependent mooring forces, which are not included in \eqref{eq:Cummins}. The UPINN makes this conceptually possible by combining ideas of universal differential equations \cite{rackauckas2021universal} and PINNs. It involves two neural networks: One network $\hat{u}$ parametrized by $\Theta_u$ for approximating the solution function $u$, and one network $\hat{g}$ parametrized by $\Theta_g$ for approximating the unknown state dynamics. As an example, consider the differential equation posed as an initial value problem (IVP) with an incomplete known right hand side function,
\begin{equation}
    \dot{u}=\mathcal{N}_{\text{known}}(u)+\mathcal{N}_\text{unknown}(u), \quad t\geq t_0,
\end{equation}
Since the solution $u(t)$ and some parts of the equation are unknown, measurement data at discrete times $u_i=u(t_i)$ is required for uniqueness. We define a loss term that minimizes the error between $\hat{u}$ and $u$ guided using available data $(t_i, u_i)$. Using the mean squared error (MSE), we can formulate this loss that accounts for data mismatch as:
\begin{equation}
    \mathcal{L}_{\text{data}}(\Theta_u)=\sum_{i=1}^{N_\text{data}}\left(\hat{u}(t_i)-u_i\right)^2.
\end{equation}
The second loss term involves the equation residual that may be describing how so-called domain knowledge, which can be calculated using AD, as in the PINN framework:
\begin{equation}
    \mathcal{L}_{\text{eq}}(\Theta_u,\Theta_g)=\sum_{i=1}^{N_\text{eq}}\left(\frac{d}{dt}{\hat{u}}(t_i)-\mathcal{N}_{\text{known}}(\hat{u})(t_i)-\hat{g}(t_i, \hat{u}(t_i))\right)^2.
\end{equation}
This loss terms serves to make $\hat{g}$ approximate $\mathcal{N}_{\text{unknown}}$. Thus, we train both surrogates $\hat{g}$ and $\hat{u}$ at the same time by minimizing some combination of these loss terms to obtain the optimal set of parameters $\Theta^*$ of the neural network by solving the optimization problem
\begin{equation}
\Theta^* = \arg \min_\Theta  \mathcal{L}(\Theta).
\end{equation}

\subsection{Estimating Hydrodynamic Coefficients}
Determining the hydrodynamic coefficients based on time series data $\{x_k(t_i)\}_{i=1}^N$ is an inverse problem. PINNs have been successfully used for this type of parameter estimation, e.g., in the original paper by Raissi et al. \cite{raissi2019physics}, where the pressure field was recovered from velocity data by using the Navier-Stokes equations to link the velocity variables to the pressure. When the forces $F_j$ in \eqref{eq:Cummins} are zero, we run into a uniqueness problem because each term on the left-hand side contains an unknown multiplicative factor, the factors being $m_{jk}+a_{jk}^\infty$, $K_{jk}(t)$, and $c_{jk}$. Hence, we cannot discover all of them at once. Knowing one of them, we can determine the others; in this paper, we assume that the hydrostatic restoring $c_{jk}$ is known.\\ \\
We are interested in identifying the dynamical system $x_k(t)$ and the hidden function $K_{jk}(t)$, so the overall model contains two surrogates, one to approximate each. We know the IRF is purely time-dependent and independent of the dynamical system, so the framework simplifies this problem. For this reason, we can train the two models independently. First, we approximate the dynamical system with a surrogate model trained on available data. If the dataset is sufficiently dense and contains little to no noise, we can compute derivatives, e.g., using finite differences, to encourage smoother gradients in the network. Velocity measurements can also be used if available. Once the system surrogate is trained, we can use it to construct the residual of the Cummins equations as in the physics-informed framework. From here, we can minimize an appropriate loss function to learn the IRF, the infinite added mass, and/or the hydrostatic restoring force. Once the IRF has been discovered, we can calculate the frequency-dependent added mass and damping coefficients using the relations \eqref{eq:IRF2A}-\eqref{eq:IRF2B}.

In this paper, we consider a one-dimensional (1D) case where only heave motion occurs, i.e., $j=k=3$, so we omit the subscripts $j$ and $k$ from now on. When rewriting \eqref{eq:Cummins} as a first-order system, we will use $x$ for the heave displacement and $v=\dot{x}$ for the heave velocity. We denote the state vector by $\boldsymbol{x}=(x,v)$.

\subsection{Neural Network Architecture}
We use the MLP architecture for the two surrogates. For the system $\boldsymbol{x}=(x,v)$, we use two decoupled networks to approximate each system variable. The total system surrogate, consisting of both the displacement and velocity networks, is denoted by $\hat{\boldsymbol{x}}=(\hat{x},\hat{v})$. Since we expect the system to show oscillating behavior, we utilize a Fourier feature embedding $\gamma(t)$ \cite{tancik2020fourier} to catch higher frequency components and mitigate spectral bias quickly, i.e., the input to the system approximators is
\begin{equation}
    \gamma(t) = \left(\cos(kBt), \sin(kBt)\right),
\end{equation}
where $B$ is a vector of trainable parameters and $k$ is a fixed tuning parameter. The system approximators for the displacement and velocity are thus
\begin{subequations}
\begin{align}
    \hat{x}(t)&=\hat{n}_x\left(\gamma_x(t)\right),\\
    \hat{v}(t)&=\hat{n}_v\left(\gamma_v(t)\right),
\end{align}
\end{subequations}
where $n_x$ and $n_v$ are MLPs. Even though we do not necessarily expect as much oscillatory behavior in the IRF, the gradients are often steep, so the Fourier feature embedding is also used in this network. Additionally, we assume the IRF goes to zero with time so that we can define the IRF surrogate as
\begin{equation}
    \hat{K}(t)=\hat{n}_K\left(\gamma_K(t)\right)\exp(-\kappa t),\label{eq:ansatz}
\end{equation}
where $\hat{n}_K$ is an MLP, and the rate $\lambda$ is a parameter we can tune based on how quickly we expect the IRF to decay. We can also let the optimizer tune $\lambda$. This technique of approximating a factor of the IRF with a neural network can facilitate faster learning since the network itself does not have to learn the decaying behavior.

\subsection{Integrating the Neural Surrogate Model}
To identify the correct IRF, we must train it to satisfy the Cummins equation. This involves the causal convolution of the neural surrogate model. This is done with a high-order accurate Gaussian quadrature rule consisting of points $\tau_i(t)$ and weights $w_i(t)$. The integration interval depends on time, so we scale the quadrature points for each time point, hence the dependence on $t$.
\begin{equation}
    \int^t_0\hat{K}(t-\tau)\hat{v}(\tau)d\tau\approx\sum_iw_i\hat{K}(t-\tau_i)\hat{v}(\tau_i).\label{eq:quadrature}
\end{equation}
To capture local behavior without excessively increasing the quadrature order, we can split the integration interval into multiple subintervals and sum their contributions; a procedure also used in the spectral element method.

This integration is generally expensive, since it involves substantial computation at each time point. This underscores the importance of using techniques that accelerate convergence, such as Fourier embedding or incorporating prior knowledge into the model.
\section{Numerical Experiments and Results}
We test the method on two simple 1D problems: A free decay of bodies such as a sphere and a box. As stated earlier, only heave motion is expected, and no other degrees of freedom are affected by the heave motion. Table \ref{tab:rhoLmac} shows relevant parameters for each case: The gravitational acceleration $g$, the water density $\rho$, the characteristic length scale $L$, the fixed mass $m$ of the structure, the true infinite added mass $a^\infty$, and the hydrostatic restoring $c$.

% \begin{table}[H]
%     \centering
%     \begin{tabular}{|l|r|r|}\hline
%         Quantity & Sphere & Box \\ \hline
%         $\rho~[\text{kg}/\text{m}^3]$ & $1000$ & $1000$ \\ \hline
%         $L~[\text{m}]$ & $10$ & $2$ \\ \hline
%         $m~[\text{kg}]$ & $261~799.4$ & $8000.011$ \\ \hline
%         $a^\infty~[\text{kg}]$ & $132~830.0$ & $3281.4168$ \\ \hline
%         $c~[\text{kg}~\text{s}^{-2}]$ & $770~475.634~2$ & $39~240.0$ \\ \hline
%     \end{tabular}
%     \caption{Values of quantities in the two test cases.}
%     \label{tab:rhoLmac}
% \end{table}

% \begin{table}[H]
%     \centering
%     \caption{Values of quantities for the two structures tested.}
%     \begin{tabular}{|c|c|c|c|c|c|c|}\hline
%         Structure & $g~[\text{m}/\text{s}^2]$ & $\rho~[\text{kg}/\text{m}^3]$ & $L~[\text{m}]$ & $m~[\text{kg}]$ & $a^\infty~[\text{kg}]$ & $c~[\text{kg}~\text{s}^{-2}]$ \\ \hline
%         Sphere & $9.81$ & $1000$ & $10$ & $261~799.4~~~$ & $132~830.0~~~~$ & $770~475.634~2$ \\ \hline
%         Box & $9.81$ & $1000$ & $~2$ & $~~~~8000.011$ & $~~~~~3281.4168$ & $39~240.0~~~~$ \\ \hline
%     \end{tabular}
%     \label{tab:rhoLmac}
% \end{table}
\begin{table}[H]
    \centering
    \caption{Values of quantities for the two structures tested.}
    \begin{tabular}{|c|c|c|c|c|c|c|}\hline
        Structure & $g~[\text{m}/\text{s}^2]$ & $\rho~[\text{kg}/\text{m}^3]$ & $L~[\text{m}]$ & $m~[\text{kg}]$ & $a^\infty~[\text{kg}]$ & $c~[\text{kg}~\text{s}^{-2}]$ \\ \hline
        Sphere & $9.81$ & $1000$ & $10$ & $2.617~994\times10^5$ & $1.328~300\times10^5$ & $7.704~756~342\times10^5$ \\ \hline
        Box & $9.81$ & $1000$ & $~2$ & $8.000~011\times10^3$ & $3.281~416~8\times10^3$ & $3.924~00\times10^4$ \\ \hline
    \end{tabular}
    \label{tab:rhoLmac}
\end{table}

%\noindent\textit{\textbf{Remark}: Scaling of Quantities}\\
Remark: the effectiveness and efficiency of the training phase depend heavily on the scales of the quantities involved. In many data science applications, datasets consisting of many features are standardized. When differential equations are involved, it is not always that simple, because the scaling cannot depend directly on the data but must be fixed. In this case, we consider a linear, homogeneous equation, so we can scale all coefficients by the same factor. The movement's values are already at reasonable scales, so we do not scale these. We scale the coefficient factors by $k_{\text{scale}}=\rho L^3$. This means that the network learns the function $K(t)/k_\text{scale}$ and the quantity $a^\infty/k_\text{scale}$ by providing $c/k_\text{scale}$. Then, we manually rescale before predicting.

\subsection{Data Generation}
The heave displacement data used for training are generated by running the DTUMotionSimulator \cite{bingham2022dtumotion} with known hydrodynamic coefficients obtained by solving spectral element discretized pseudo-impulsive initial boundary value problems \cite{visbech2024solving}.

We compute fourth-order finite-difference approximations of the velocity and acceleration from the displacement data. The total dataset consisting of $N$ data points is thus
\begin{equation}
    \left\{(t_i, x_i, v_i, a_i)\right\}_{i=1}^N.\label{eq:dataset}
\end{equation}
In these experiments, the data points are equally spaced in time, ranging from $0.01~\text{s}$ to $30~\text{s}$. We consider three cases in terms of data quality:
\begin{enumerate}
    \item In this case, we use all data points in the interval, hence $N=3000$.
    \item In this case, we use all data points again, i.e., $N=3000$, but we add $5\%$ relative noise to $x_i,v_i,a_i$ (after the finite difference calculations).
    \item In this case, we use only every 50th data point, so $N=60$ and we have two data points per second, i.e. $t_1=0.5~\text{s},t_2=1.0~\text{s},\dots,t_{60}=30.0~\text{s}$.
\end{enumerate}

When learning the hydrodynamic coefficients, we can sample the time domain rather than rely on specific measurements. We do this with a Sobol sequence scaled to the interval $(0~\text{s}, 30~\text{s})$. In all three cases above, we use $N_t=3000$ sample points for this training phase. In every case, the floating-point precision of all data points and parameters is 64-bit.

\subsection{Approximating the System}
The total set of parameters in each network of $\hat{\boldsymbol{x}}$ is denoted by $\Theta_x$ and $\Theta_v$, respectively. These networks are trained on the dataset \eqref{eq:dataset}. The loss terms for the system are constructed as:
% \begin{subequations}
% \begin{align}
%     \mathcal{L}_1(\Theta_x)&=\sum_{i=1}^{N}\left(\hat{x}(t_i)-x_i\right)^2,\\
%     \mathcal{L}_2(\Theta_x, \Theta_v)&=\sum_{i=1}^{N}\left(\frac{d}{dt}\hat{x}(t_i)-\hat{v}(t_i)\right)^2,\\
%     \mathcal{L}_3(\Theta_v)&=\sum_{i=1}^{N}\left(\hat{v}(t_i)-v_i\right)^2,\\
%     \mathcal{L}_4(\Theta_v)&=\sum_{i=1}^{N}\left(\frac{d}{dt}\hat{v}(t_i)-a_i\right)^2.
% \end{align}
% \label{eq:data_res}
% \end{subequations}
\begin{equation}
    \mathcal{L_\text{sys}}(\Theta_x,\Theta_v)=\lambda_1\sum_{i=1}^{N}\left(\hat{x}(t_i)-x_i\right)^2+\lambda_2\sum_{i=1}^{N}\left(\frac{d}{dt}\hat{x}(t_i)-\hat{v}(t_i)\right)^2+\lambda_3\sum_{i=1}^{N}\left(\hat{v}(t_i)-v_i\right)^2+\lambda_4\sum_{i=1}^{N}\left(\frac{d}{dt}\hat{v}(t_i)-a_i\right)^2.\label{eq:loss_sys}
\end{equation}
Even though only the first two loss terms are necessary for the problem to be well-defined, we find that the last two terms help the model converge faster and slightly increase its accuracy. If the available data set is too sparse or noisy, these terms can be omitted, as they might introduce even larger errors.

% As mentioned, the system approximation does not depend on any other unknowns, so we can train it before approximating the coefficients of real interest. Once trained, we can freeze the parameters of $\hat{\boldsymbol{x}}$ and solve the inverse problem.

% \subsubsection*{Architecture and Optimiser}
For both networks $\hat{x}$ and $\hat{v}$, we use two hidden layers of 32 neurons and hyperbolic tangent activation. Since we expect an oscillatory solution, we use a Fourier feature embedding of the input with 32 features and $k=1$ for each network.

\begin{figure}[h]
    \centering
    \begin{subfigure}{0.48\textwidth}
        \centering
        \includegraphics[width=1\linewidth, trim=0 10cm 0 0, clip]{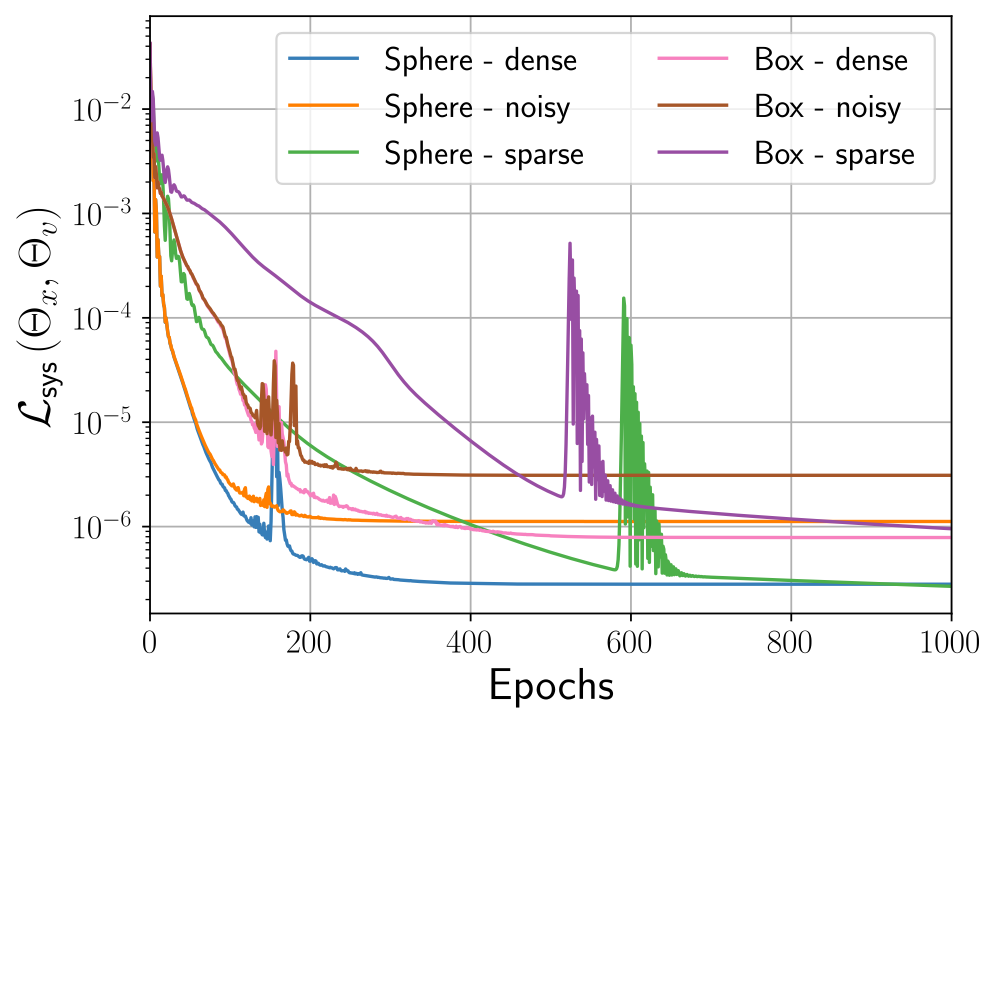}
        \caption{}
        \label{fig:data_loss_a}
    \end{subfigure}
    \hfill
    \begin{subfigure}{0.48\textwidth}
        \centering
        \includegraphics[width=\linewidth, trim=0 10cm 0 0, clip]{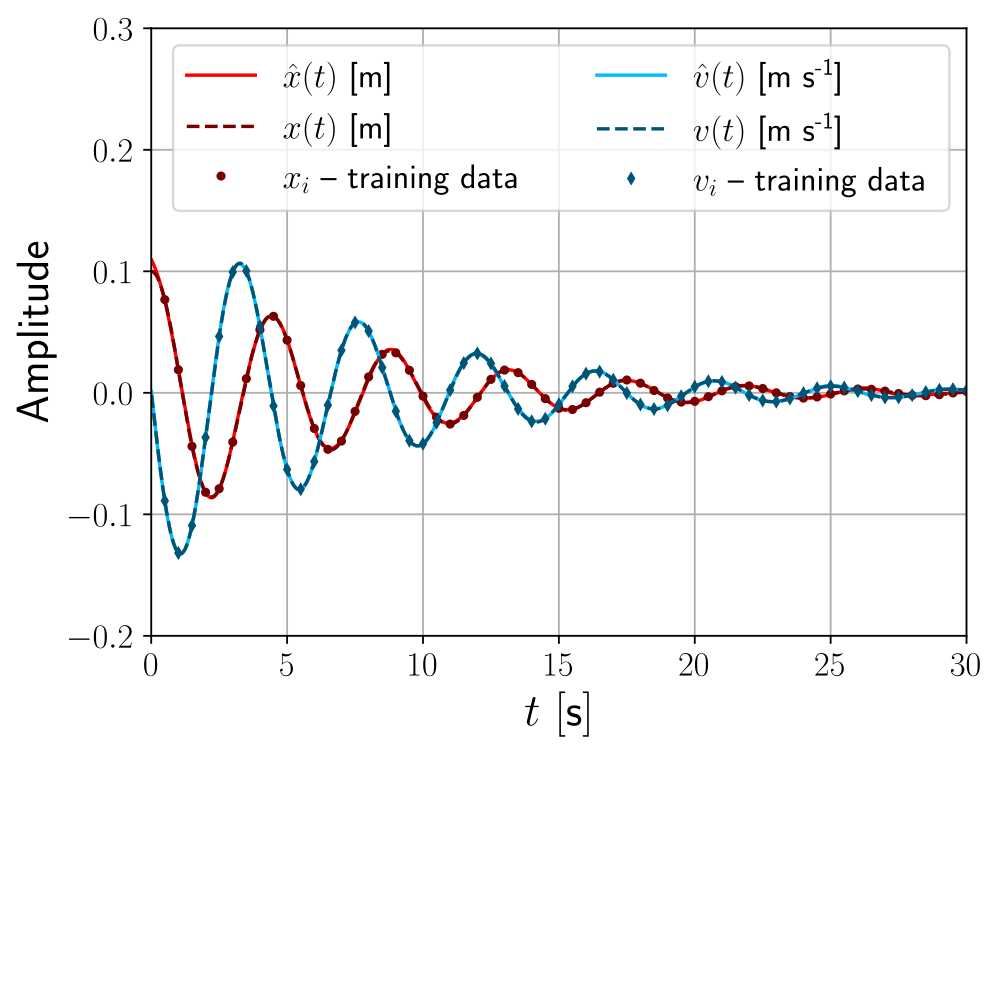}
        \caption{}
    \end{subfigure}
    \caption{(a) Loss history for the system in the three cases, i.e., based on dense, noisy, and sparse data, for a sphere and a box. (b) Example of the fitted system in the case of a sphere and sparse data, corresponding to the green line in Fig.~\ref{fig:data_loss_a}.}
    \label{fig:data_loss}
\end{figure}

We train the networks using the Adam optimizer \cite{kingma2017adam} for 1000 epochs. For the Adam optimizer, we use an initial learning rate of 0.005 and an exponential decay rate of 0.9995. All loss weights $\lambda_k$ in \eqref{eq:loss_sys} were set to $0.25$. The loss history can be seen in Fig.~\ref{fig:data_loss} a) and the resulting neural solutions in Fig.~\ref{fig:data_loss} b) obtained after the completed training. The results here demonstrate an accurate fit to the training data and a good reproduction of the system dynamics, as reflected in the body movement's position and velocity curves.

% \begin{figure}[h]
% \centering
%     \begin{subfigure}{0.4\textwidth}
%         \centering
%         \includegraphics[width=0.5\linewidth]{missing_figure.png}
%         \caption{Sphere in free decay.}
%     \end{subfigure}
%     \hfill
%     \begin{subfigure}{0.4\textwidth}
%         \centering
%         \includegraphics[width=0.5\linewidth]{missing_figure.png}
%         \caption{Box in free decay.}
%     \end{subfigure}
%     \caption{The system approximation $(\hat{x},\hat{v})$ compared to the data. The displacement data were available while the velocity data were calculated from the displacement using a fourth-order finite difference scheme. All four of the residuals in [REF] were used in training.}
%     % \label{fig:enter-label}
% \end{figure}

\subsection{Recovering Approximate Values of the Hydrodynamic Coefficients}
With a good approximation of the system, we solve for the hydrodynamic coefficients in \eqref{eq:Cummins}. We assume the hydrostatic restoring $c$ is known and denote the trainable parameter for the infinite added mass by $\hat{a}^\infty$. The trainable parameters for this problem are thus the parameters $\Theta_K$ of the surrogate $\hat{K}$ in addition to $\hat{a}^\infty$. The decay rate $\kappa$ is set to $0.5$ and fixed during training.

The residual for the Cummins equation is:
\begin{equation}
    \mathcal{R}_\text{eq}(t)=(m+\hat{a}^\infty)\frac{d}{dt}\hat{v}(t)+\int_0^t\hat{K}(t-\tau)\hat{v}(\tau)d\tau+c\hat{x}(t),\label{eq:cummins1d_res}
\end{equation}
where the force term is left out since it is zero in the case of free decay.
% This is just one of the equations in the first-order system formulation. The other equation,
% \begin{equation}
%     \frac{d}{dt}\hat{x}(t)=\hat{v}(t),
% \end{equation}
% has already been satisfied in the training phase.

% The external forces are zero in the case of free decay. Assuming the hydrostatic restoring $c$ is known, we want to estimate the infinite added mass $\hat{a}^\infty$ and the IRF $\hat{K}(t)$.

Additionally, to emphasize the asymptotic behavior of the added mass and damping coefficients, we define the residuals,
\begin{subequations}
\begin{align}
    \mathcal{R}_{A}(t)&=\frac{1}{\omega^\infty}\int_0^{t^\infty} \hat{K}(t)\sin(\omega^\infty t)dt,\\
    \mathcal{R}_{B}(t)&=\int_0^{t^\infty} \hat{K}(t)\cos(\omega^\infty t)dt,
\end{align}
\end{subequations}
where $\omega^\infty$ is the largest relevant frequency, which we expect to be close to 0. The upper limit $t^\infty$ of the integrals is set to the largest value of $t$ for which we have data; in this case, $t^\infty=30~\text{s}$. We add one last residual to the problem, namely a Laplace-transformed version of \eqref{eq:cummins1d_res}.
\begin{equation}
    \mathcal{R}_{\mathrm{Lap}}(s)=\mathfrak{L}\left\{(m+\hat{a}^\infty)\frac{d}{dt}\hat{v}+c\hat{x}\right\}(s)+\mathfrak{L}\left\{\hat{K}\right\}(s)\cdot\mathfrak{L}\left\{\hat{v}\right\}(s),\label{eq:cummins1d_res_laplace}
\end{equation}
where $\mathfrak{L}$ denotes the Laplace transform. This equation utilizes the properties of the convolution theorem. The cosine, sine, and Laplace transforms are calculated using a quadrature rule similar to the convolution integral in \eqref{eq:quadrature}. For the Laplace loss, the variable $s$ is randomly sampled in each iteration. The loss function for the inverse problem is thus
% The total loss function to be minimised is defined by
% \begin{equation}
%     \mathcal{C}=\lambda_{\mathrm{eq}}\mathrm{MSE}\left(\mathcal{R}_{\mathrm{eq}}\right)+\lambda_{A}\mathrm{MSE}\left(\mathcal{R}_{A}\right)+\lambda_{B}\mathrm{MSE}\left(\mathcal{R}_{B}\right)+\lambda_{\mathrm{Laplace}}\mathrm{MSE}\left(\mathcal{R}_{\mathrm{Laplace}}\right).\label{eq:cummins_loss}
% \end{equation}

% \begin{subequations}
% \begin{align}
%     \mathcal{L}_\text{eq}(\Theta_K,\hat{a}^\infty)&=\sum_{i=1}^{N_{t}}\left(\mathcal{R}_\text{eq}(t_i)\right)^2,\label{eq:inv_loss_first}\\
%     \mathcal{L}_\mathcal{A}(\Theta_K)&=\left(\mathcal{R}_\mathcal{A}(t^\infty)\right)^2,\\
%     \mathcal{L}_\mathcal{B}(\Theta_K)&=\left(\mathcal{R}_\mathcal{B}(t^\infty)\right)^2,\\
%     \mathcal{L}_\text{Lap}(\Theta_K,\hat{a}^\infty)&=\sum_{i=1}^{N_{s}}\left(\mathcal{R}_\text{Lap}(t_i)\right)^2.\label{eq:inv_loss_last}
% \end{align}
% \end{subequations}

\begin{equation}
    \mathcal{L}_\text{coeff}(\Theta_K,\hat{a}^\infty)=\lambda_1\sum_{i=1}^{N_{t}}\left(\mathcal{R}_\text{eq}(t_i)\right)^2+\lambda_2\left(\mathcal{R}_\mathcal{A}(t^\infty)\right)^2+\lambda_3\left(\mathcal{R}_\mathcal{B}(t^\infty)\right)^2+\lambda_4\sum_{i=1}^{N_{s}}\left(\mathcal{R}_\text{Lap}(t_i)\right)^2.\label{eq:loss_irf}
\end{equation}

\subsubsection{Neural Network Architecture and Optimizer}
We minimize the loss function \eqref{eq:loss_irf} with respect to $\Theta_K$ and $\hat{a}^\infty$. The parameter $\hat{a}^\infty$ is initialized with the same value as the known mass $m$. The system surrogate $\hat{\boldsymbol{x}}$ is already trained, so its parameters are fixed in the inverse problem. We represent the IRF by the surrogate defined in \eqref{eq:ansatz}. For the neural network, we use a simple MLP with two hidden layers of 64 neurons and a Fourier embedding with $k=1$ and 64 features. We use the same optimizer and settings as for system training. The loss terms had fixed weightings of $ 5, 1, 50, 5$, normalized so that they sum to 1. Additionally, we used self-adaptive weights for the data points in the first loss term.

\begin{figure}[H]
    \centering
    \begin{subfigure}{0.48\textwidth}
        \centering
        \includegraphics[width=\linewidth, trim=0 10cm 0 0, clip]{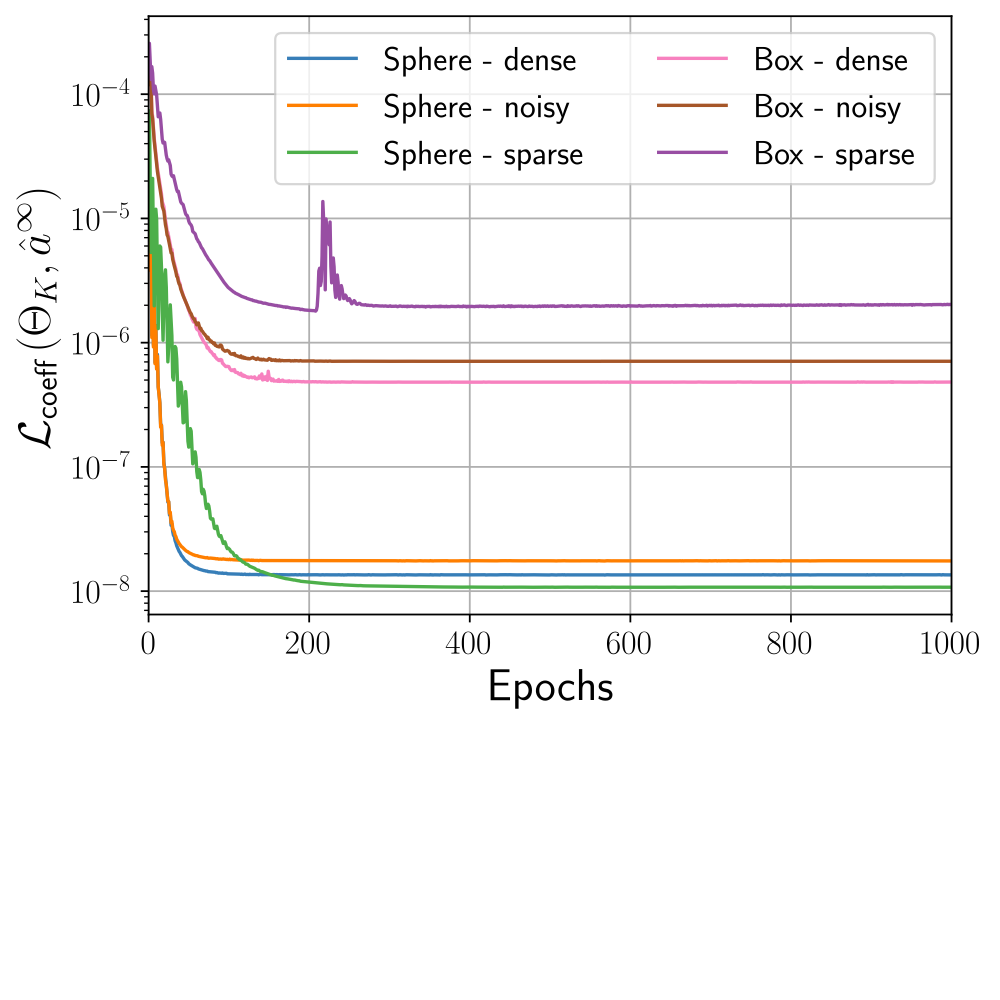}
        \caption{}
        \label{fig:coeff_loss_a}
    \end{subfigure}
    \hfill
    \begin{subfigure}{0.48\textwidth}
        \centering
        \includegraphics[width=\linewidth, trim=0 10cm 0 0, clip]{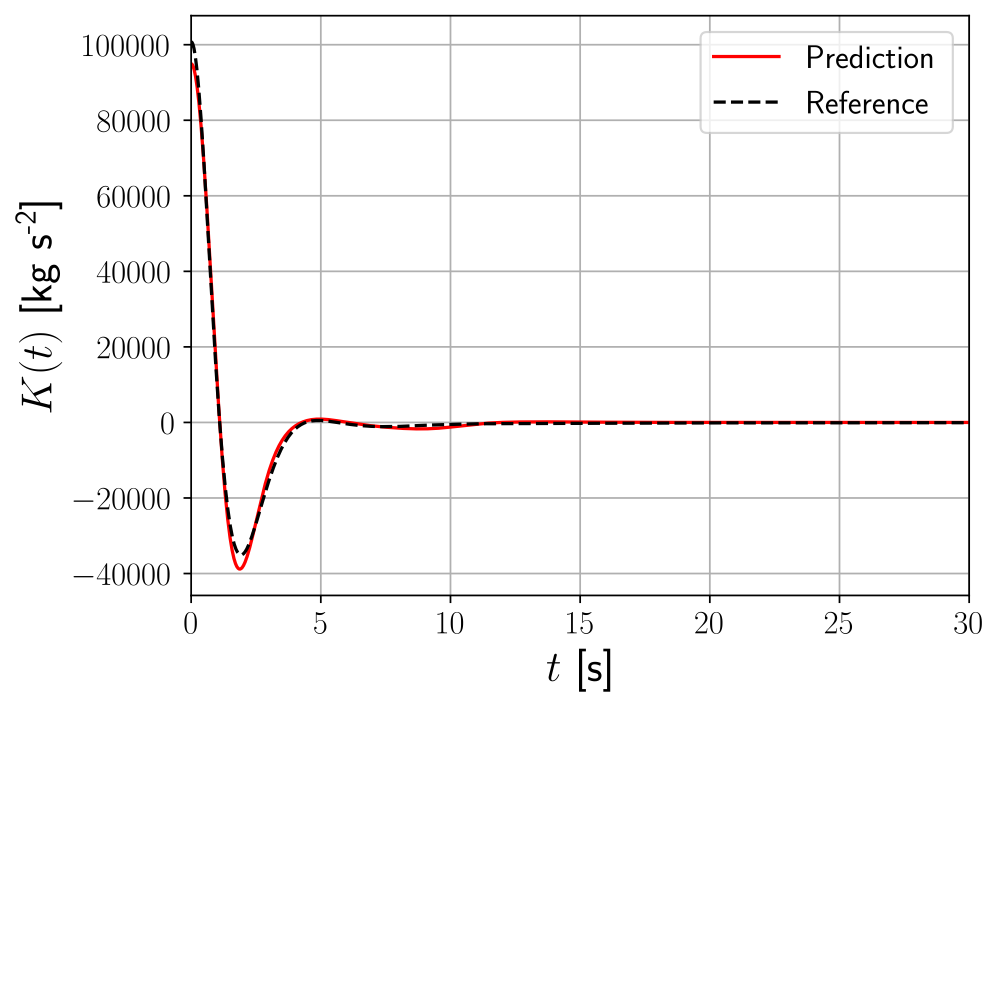}
        \caption{}
    \end{subfigure}
    \caption{(a)~Loss history for the inverse problem of estimating the IRF $K(t)$ and the infinite added mass $a^\infty$. (b)~Example of fitted IRF in the case of a sphere and sparse data, corresponding to the green line in Fig.~\ref{fig:coeff_loss_a}.}
    \label{fig:inverse_loss}
\end{figure}

Figure~\ref{fig:inverse_loss} shows the loss history for the inverse problem and the estimated IRF for the sphere in the case of sparse data. Table \ref{tab:inf_added_mass_results} shows estimations of the infinite added mass for the sphere and box in all three cases.

\begin{figure}[H]
    \centering
    \includegraphics[width=0.48\linewidth, trim=0 10cm 0 0, clip]{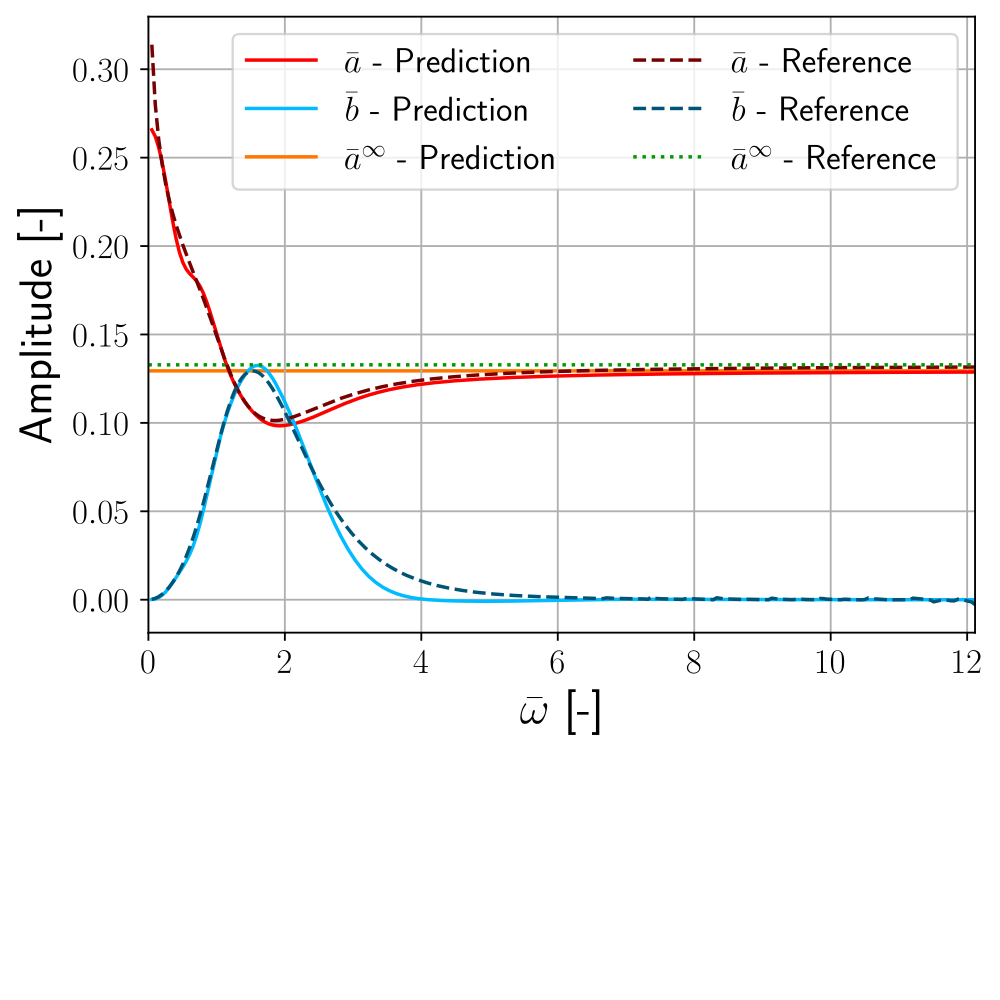}
    \caption{Estimation of the frequency-dependent added mass and damping coefficients for the sphere in the case of sparse data. All quantities have been normalized to dimensionless quantities according to the WAMIT user manual.}
    \label{fig:coeff_estimate}
\end{figure}

When the IRF surrogate has been trained, the added mass and damping are calculated according to \eqref{eq:IRF2A}-\eqref{eq:IRF2B}. Figure \ref{fig:coeff_estimate} shows the estimates for the sphere, again in the case of sparse data. The coefficients have been normalized to dimensionless quantities, as in \cite{wamit}.
% Small networks seem to consistently give better results. This is most likely due to the fact that the loss is defined through the integral of $\hat{K}$, and so, a model that is too flexible will be more unstable during training. We also use very few Fourier features postulating that the IRF can be represented as a function of only a handful of frequencies.

% This also means that we reduce the number of total Fourier features to only 16 compared to the 32 in the other networks.

% \begin{table}[]
%     \centering
%     \begin{tabular}{|l|c|c|}\hline
%          & Sphere & Box  \\ \hline
%          Reference & $1$ & $1$ \\ \hline
%          Initial & $1$ & $1$ \\ \hline
%          Case 1 & $1$ & $1$ \\ 
%          Rel. err. & $1$ & $1$ \\ \hline
%          Case 2 & $1$ & $1$ \\ 
%          Rel. err. & $1$ & $1$ \\ \hline
%          Case 3 & $1$ & $1$ \\ 
%          Rel. err. & $1$ & $1$ \\ \hline
%     \end{tabular}
%     \caption{Infinite added mass $a^\infty$: True value, initial value (before training) and learned values for the three cases.}
%     \label{tab:inf_added_mass_results}
% \end{table}

{\footnotesize
\begin{table}[H]
    \centering
    \caption{Infinite added mass $a^\infty$: True value, initial value (before training), and learned values for the three cases.}
    \begin{tabular*}{\textwidth}{@{\extracolsep{\fill}}|c|c|c|c|c|c|c|c|} \hline
        Structure & -  & \multicolumn{2}{c|}{Case 1: Dense} & \multicolumn{2}{c|}{Case 2: Noisy} & \multicolumn{2}{c|}{Case 3: Sparse} \\ \hline
        Type & Reference & Pred. & Rel. err. & Pred. & Rel. err.  & Pred. & Rel. err. \\ 
        & [kg] & [kg] & [\%] & [kg] & [\%] & [kg] & [\%]  \\ \hline
        Sphere & 1.328e-01 & 1.309e-01 & -1.47 & 1.346e-01 & 1.30 & 1.294e-01 & -2.56  \\ \hline
        Box & 4.102e-01 & 4.058e-01 & -1.05 & 4.215e-01& 2.76 & 5.585e-01 & 36.1 \\ \hline
    \end{tabular*}
    \label{tab:inf_added_mass_results}
\end{table}
}

\subsection{Discussion}
The accuracy of the hydrodynamic coefficients depends on the accuracy of the learned movement DOFs, which in turn rely on accurate measurements. Thus, multiple steps in this method can amplify the final approximation errors. Noise propagation from measurements to estimated coefficients cannot be entirely avoided. Still, it can be minimized by avoiding overfitting and, where possible, incorporating known behaviors into the surrogates. The coefficient estimates from noisy data were less accurate than those from dense, noiseless data, but the difference was insignificant. In the case of sparse data, however, the box movement might be too high-frequency for the method to obtain reasonable estimates. This is shown in Tab. \ref{tab:inf_added_mass_results} where the infinite added mass estimate is much worse in this case than in any other due to the sparse (limited) set of data to learn from.

\section{Conclusions}
This paper has presented results for estimating hydrodynamic coefficients obtained by solving an inverse problem involving the Cummins equations with physics-informed neural networks (PINNs). We explained how to use universal PINNs (UPINNs) for this purpose when state-dependent forces are unknown, although this case was not included in the experiments. The estimation method was tested using data from the free decay of a sphere and a box. The model provides fair estimates of the coefficients, depending on the quality of the available movement data, and it was noted that accurate estimates could be obtained. Specifically, the results highlight the importance of the data density for movement frequencies. 

A current limitation of the work is that the proposed idea is based on time-series data generated by numerical simulations to verify the concept of using physics-informed neural networks to estimate hydrodynamic coefficients for floating offshore structures from movement data. In ongoing work, we will explore the idea in the context of measured time-series data from sensors that may also exhibit noise.

\section{Acknowledgments}
The research was conducted at the Technical University of Denmark (DTU) in the Department of Applied Mathematics and Computer Science with computational resources provided by the DTU Computing Center (DCC). This work was supported by COWIfonden under Grant No. A-165.19 in the research project: ‘‘A new digital twin concept for floating offshore structures’’.

\bibliographystyle{abbrvnat}
\bibliography{references.bib}  

@software{bingham2022dtumotion,
    author = {Bingham, H. B. and Read, R.},
    year = {(2022)},
    title = {DTUMotionSimulator: A Matlab Package for Simulating Linear or Weakly Nonlinear Response of a Floating Structure to Ocean Waves},
    organisation = {Technical University of Denmmark}
}

@article{cao2024deep,
title = {Deep neural operators can predict the real-time response of floating offshore structures under irregular waves},
journal = {Computers \& Structures},
volume = {291},
pages = {107228},
year = {2024},
author = {Qianying Cao and Somdatta Goswami and Tapas Tripura and Souvik Chakraborty and George Em Karniadakis}
}

@article{Rumelhart1986LearningErrors,
    title = {{Learning representations by back-propagating errors}},
    year = {1986},
    journal = {Nature},
    author = {Rumelhart, David E. and Hinton, Geoffrey E. and Williams, Ronald J.},
    number = {6088},
    volume = {323},
}

@article{lee1990neural,
  title={Neural algorithm for solving differential equations},
  author={Lee, Hyuk and Kang, In Seok},
  journal={Journal of Computational Physics},
  volume={91},
  number={1},
  pages={110--131},
  year={1990},
  month={November}
}

@article{psichogios1992hybrid,
  title={A hybrid neural network-first principles approach to process modeling},
  author={Psichogios, D. C. and Ungar, L. H.},
  journal={AIChE Journal},
  volume={38},
  pages={1499--1511},
  year={1992}
}

@article{dissanayake1994neural,
  title={Neural-network-based approximations for solving partial differential equations},
  author={Dissanayake, M. W. M. G. and Phan-Thien, Nhan},
  journal={Communications in Numerical Methods in Engineering},
  volume={10},
  number={3},
  pages={195--201},
  year={1994}
}

@article{lu2021deepxde,
  title={DeepXDE: A deep learning library for solving differential equations},
  author={Lu, Lu and Meng, Xuhui and Mao, Zhiping and Karniadakis, George Em},
  journal={SIAM Review},
  volume={63},
  number={1},
  pages={208--228},
  year={2021}
}

@techreport{baker2019workshop,
  title={Workshop Report on Basic Research Needs for Scientific Machine Learning: Core Technologies for Artificial Intelligence},
  author={Baker, N. and Alexander, F. and Bremer, T. and Hagberg, A. and Kevrekidis, Y. and Najm, H. and Parashar, M. and Patra, A. and Sethian, J. and Wild, S. and others},
  institution={U.S. DOE Office of Science},
  address={Washington, DC},
  year={2019}
}

@article{DiColaEtAl2022,
	author = {Cuomo, Salvatore and Di Cola, Vincenzo Schiano and Giampaolo, Fabio and Rozza, Gianluigi and Raissi, Maziar and Piccialli, Francesco},
	journal = {Journal of Scientific Computing},
	number = {3},
	pages = {88},
	title = {Scientific Machine Learning Through Physics--Informed Neural Networks: Where we are and What's Next},
	volume = {92},
	year = {2022},
}

@inproceedings{MortensenEtAl2021,
    author = {Mortensen, Line K. and Laskowski, Wojciech and Engsig-Karup, Allan P. and Eskilsson, Claes and Monteserin, Carlos},
    title = {Simulation of NonlinearWaves Interacting with a Heaving Body using a p-Multigrid Spectral Element Method},
    volume = {The 31st International Ocean and Polar Engineering Conference},
    series = {International Ocean and Polar Engineering Conference},
    pages = {ISOPE-I-21-3137},
    year = {2021},
    month = {06},
}

@book{cummins1962impulse,
  title={The Impulse Response Function and Ship Motions},
  author={Cummins, W.E. and David W. Taylor Model Basin},
  series={Report (David W. Taylor Model Basin)},
  year={1962},
  publisher={Department of the Navy, David Taylor Model Basin},
}

@article{cybenko1989approximation,
  author = {Cybenko, G.},
  title = {Approximation by superpositions of a sigmoidal function},
  language = {eng},
  format = {article},
  journal = {Mathematics of Control, Signals and Systems},
  volume = {2},
  number = {4},
  pages = {303-314},
  year = {1989},
  publisher = {Springer-Verlag},
}

@article{hornik1989multilayer,
    title = {Multilayer feedforward networks are universal approximators},
    journal = {Neural Networks},
    volume = {2},
    number = {5},
    pages = {359-366},
    year = {1989},
    author = {Kurt Hornik and Maxwell Stinchcombe and Halbert White},
}

@misc{kingma2017adam,
      title={Adam: A Method for Stochastic Optimization}, 
      author={Diederik P. Kingma and Jimmy Ba},
      year={2017},
      eprint={1412.6980},
      archivePrefix={arXiv},
      primaryClass={cs.LG},
}

@article{lagaris1998artificial,
  author = {Lagaris, Isaac Elias and Likas, Aristidis and Fotiadis, Dimitrios I.},
  title = {Artificial neural networks for solving ordinary and partial differential equations},
  language = {eng},
  format = {article},
  journal = {Ieee Transactions on Neural Networks},
  volume = {9},
  number = {5},
  pages = {987-1000},
  year = {1998},
}

@phdthesis{liapis1986time,
    author = {Liapis, Stergios John},
    title = {Time-domain analysis of ship motions},
    school = {University of Michigan},
    year = {1986}
}

@article{mclenny2023selfadaptive,
title = {Self-adaptive physics-informed neural networks},
journal = {Journal of Computational Physics},
volume = {474},
pages = {111722},
year = {2023},
author = {Levi D. McClenny and Ulisses M. Braga-Neto},
}

@article{KURNIA2023108885,
title = {NEMOH: Open-source boundary element solver for computation of first- and second-order hydrodynamic loads in the frequency domain},
journal = {Computer Physics Communications},
volume = {292},
pages = {108885},
year = {2023},
author = {Ruddy Kurnia and Guillaume Ducrozet},
}

@incollection{paszke2019pytorch,
title = {PyTorch: An Imperative Style, High-Performance Deep Learning Library},
author = {Paszke, Adam and Gross, Sam and Massa, Francisco and Lerer, Adam and Bradbury, James and Chanan, Gregory and Killeen, Trevor and Lin, Zeming and Gimelshein, Natalia and Antiga, Luca and Desmaison, Alban and Kopf, Andreas and Yang, Edward and DeVito, Zachary and Raison, Martin and Tejani, Alykhan and Chilamkurthy, Sasank and Steiner, Benoit and Fang, Lu and Bai, Junjie and Chintala, Soumith},
booktitle = {Advances in Neural Information Processing Systems 32},
pages = {8024--8035},
year = {2019},
publisher = {Curran Associates, Inc.},
}

@article{podina2023universal,
  author = {Podina, L. and Eastman, B. and Kohandel, M.},
  title = {Universal Physics-Informed Neural Networks: Symbolic Differential Operator Discovery with Sparse Data},
  format = {article},
  journal = {Proc. Mach. Learn. Res.},
  volume = {202},
  pages = {27948-27956},
  year = {2023},
  publisher = {ML Research Press}
}

@article{rackauckas2021universal,
  author = {Rackauckas, Christopher and Ma, Yingbo and Martensen, Julius and Warner, Collin and Zubov, Kirill and Supekar, Rohit and Skinner, Dominic and Ramadhan, Ali and Edelman, Alan},
  title = {Universal Differential Equations for Scientific Machine Learning},
  language = {und},
  format = {article},
  pages = {29},
  year = {2021}
}

@article{raissi2019physics,
  author = {Raissi, M. and Perdikaris, P. and Karniadakis, G. E.},
  title = {Physics-informed neural networks: A deep learning framework for solving forward and inverse problems involving nonlinear partial differential equations},
  format = {article},
  journal = {J. Comput. Phys.},
  volume = {378},
  pages = {686-707},
  year = {2019},
  publisher = {Academic Press Inc.}
}

@article{tancik2020fourier,
  author = {Tancik, Matthew and Srinivasan, Pratul P. and Mildenhall, Ben and Fridovich-Keil, Sara and Raghavan, Nithin and Singhal, Utkarsh and Ramamoorthi, Ravi and Barron, Jonathan T. and Ng, Ren},
  title = {Fourier features let networks learn high frequency functions in low dimensional domains},
  language = {eng},
  format = {article},
  journal = {Advances in Neural Information Processing Systems},
  volume = {2020-},
  year = {2020},
  publisher = {Neural information processing systems foundation}
}

@article{visbech2024solving,
  author = {Visbech, J. and Engsig-Karup, A. P. and Bingham, H. B.},
  title = {Solving the complete pseudo-impulsive radiation and diffraction problem using a spectral element method},
  format = {article},
  journal = {Comput. Methods Appl. Mech. Eng.},
  volume = {423},
  pages = {116871},
  year = {2024},
  publisher = {Elsevier B.V.},
}

@software{wamit,
  author = {{WAMIT Inc.}},
  title = {{WAMIT}},
  url = {https://www.wamit.com},
  version = {7.5},
  date = {2023},
}

@article{Karniadakis2021,
  title   = {Physics-informed machine learning},
  author  = {Karniadakis, George Em and Kevrekidis, Ioannis G. and Lu, Lu and Perdikaris, Paris and Wang, Sifan and Yang, Liu},
  journal = {Nature Reviews Physics},
  volume  = {3},
  number  = {6},
  pages   = {422--440},
  year    = {2021},
}
% \printbibliography
\end{document}